\documentclass[12pt,a4paper,twoside]{scrartcl}
\usepackage{graphicx}
\usepackage{amsmath}
\usepackage{amssymb}
\usepackage[figuresright]{rotating}
\usepackage{fancyhdr}
\usepackage[width=16cm]{geometry} 
\newcommand{\TheAuthor}{}
\renewcommand{\TheAuthor}{ A. N. Ivanov {\it et al.}}

\def\skiplinehalf{\medskip\\}

\def\supit#1{\raisebox{0.8ex}{\small\it #1}\hspace{0.05em}}  

\title{Recent Theoretical Studies on \\Hadronic Atoms}
{\author{{\bf A. N. Ivanov}\supit{a}\supit{b}\supit{c}, M.
    Cargnelli\supit{b}, M. Faber\supit{a}, H. Fuhrmann\supit{b},\\ V.
    A.  Ivanova\supit{c}, J. Marton\supit{b}, N. I.
    Troitskaya\supit{c}, J. Zmeskal\supit{b} \skiplinehalf
    \supit{a}Atomic Institute, Vienna
    University of Technology, Vienna, Austria \\
    \supit{b}Stefan Meyer Institute, Austrian
    Academy of Sciences, Vienna, Austria \\
    \supit{c}State Polytechnic University of St. Petersburg, St.
    Petersburg, Russia}

\date{}
\begin{document}

\maketitle
\begin{abstract}
  A report contains a review of recent theoretical investigations on
  kaonic atoms carried out at Stefan Meyer Institute of the Austrian
  Academy of Sciences. We discuss (1) a phenomenological quantum field
  theoretic model for $\bar{K}N$ interactions at threshold of the
  reactions $K^-p \to K^-p$, $K^-n \to K^-n$ and $K^-d \to K^-d$, and
  (2) the energy level displacements of the ground and $nP$ excited
  states of kaonic hydrogen, (3) the contribution of the $\sigma^{I =
    1}_{KN}(0)$--term to the shift of the energy level of the ground
  state of kaonic hydrogen, (4) the isospin--breaking and dispersive
  corrections to the energy level displacement of the ground state of
  kaonic hydrogen, (5) the radiative transitions $nP \to 1S + \gamma$,
  induced by strong low--energy interactions and enhanced by the
  Coulomb interaction, in kaonic hydrogen and kaonic deuterium, (6)
  Perspectives and (7) Comments on the approach, where we adduce our
  recent results obtained after the Workshop EXA05.
\end{abstract}
\section{Kaonic hydrogen}
\setcounter{equation}{0}
\subsection{Experimental data on the energy level displacement of the
  ground state of kaonic hydrogen}

The contemporary experimental and theoretical status of kaonic atoms
has been recently outlined by J\"urg Gasser \cite{JG04}.  The most
recent experimental value on the energy level displacement of the
ground state of kaonic hydrogen
\begin{eqnarray}\label{label1}
  - \epsilon^{(\exp)}_{1s} + i\,\frac{\Gamma^{(\exp)}_{1s}}{2} = 
(- 194 \pm 41) + i\,(125 \pm 59)\,{\rm eV}, 
\end{eqnarray} 
obtained by the DEAR Collaboration \cite{DEAR}, by a factor 2 smaller
than the experimental value measured by the KEK Collaboration 
\cite{KEK}
\begin{eqnarray}\label{label2}
  - \epsilon^{(\exp)}_{1s} + i\,\frac{\Gamma^{(\exp)}_{1s}}{2} = 
(- 323 \pm 64) + i\,(204\pm 115)\,{\rm eV}.
\end{eqnarray} 
For the theoretical analysis of the experimental data on the energy
level displacement of the ground state of kaonic hydrogen, obtained by
the DEAR Collaboration Eq.(\ref{label1}), we have proposed a
phenomenological quantum field theoretic model of strong low--energy
$\bar{K}N$ interactions at threshold \cite{IV3}.

\subsection{Phenomenological quantum field theoretic model for
  $\bar{K}N$ interactions at threshold}

In Ref.\cite{IV3} (see also \cite{IV4}) we have proposed a
phenomenological quantum field theoretic model for strong low--energy
$\bar{K}N$ interactions at threshold and calculated the complex S--wave
scattering lengths $\tilde{a}^0_0$ and $\tilde{a}^1_0$ of $\bar{K}N$
scattering with isospin $I = 0$ and $I = 1$:
\begin{eqnarray}\label{label3}
  \tilde{a}^0_0 &=& (-\,1.221 \pm 0.072) + i\,(0.537 \pm 0.064)\,{\rm fm},
\nonumber\\ 
  \tilde{a}^1_0 &=& (+\,0.258
  \pm 0.024) + i\,(0.001 \pm 0.000) \,{\rm fm}.
\end{eqnarray}
The S--wave amplitude of $K^-p$ scattering at threshold is equal to \cite{IV3}:
\begin{eqnarray}\label{label4}
  f^{K^-p}_0(0) &=& \frac{\tilde{a}^0_0 + \tilde{a}^1_0}{2} = 
{\cal R}e\,f^{K^-p}_0(0) + i\,{\cal
    I}m\,f^{K^-p}_0(0) =\nonumber\\
  &=&  (-\,0.482 \pm 0.034) + i\,(0.269 \pm
  0.032)\,{\rm fm}
\end{eqnarray}
The theoretical value for the energy level displacement of the ground
state of kaonic hydrogen  is defined by the DGBTT formula (the
Deser--Goldberger--Baumann--Thirring--Trueman formula \cite{DT54}):
\begin{eqnarray}\label{label5}
  - \,\epsilon^{(0)}_{1s} + i\,\frac{\Gamma^{(0)}_{1s}}{2} &=& 
  2\,\alpha^3\mu^2 
  \,f^{K^-p}_0(0) =  412\,[(-\,0.482 \pm 0.034) + i\,(0.269 \pm
  0.032)] = \nonumber\\
  &=& (- 203 \pm 15) + i\,(113 \pm 14)\,{\rm eV}. 
\end{eqnarray}
Here $2\alpha^3\mu^2 = 421\,{\rm eV}/{\rm fm}$, where $\mu = m_K
m_p/(m_K + m_p) = 324\,{\rm MeV}$ is the reduced mass of the $K^-p$
pair, calculated for $m_K = 494\,{\rm MeV}$ and $m_p = 940\,{\rm MeV}$,
and $\alpha = 1/137.036$ is the fine--structure constant \cite{PDG04}.

  The theoretical values of the energy level shift and width
  Eq.(\ref{label4}) agree well with the experimental data by the DEAR
  Collaboration \cite{DEAR} Eq.(\ref{label1}) and only qualitatively
  with the experimental data by the KEK Collaboration \cite{KEK}.

  The theoretical ground for our model of low--energy $\bar{K}N$
  interactions is a phenomenological Effective Field Theory 
  with Effective Chiral Lagrangians and Chiral Perturbation Theory
  (ChPT) by Gasser and Leutwyler \cite{JG83}--\cite{UM01}. 

  In our model of strong low--energy $\bar{K}N$ interactions near
  threshold proposed in \cite{IV3,IV4} the imaginary part of the
  S--wave amplitude of $K^-p$ scattering is defined by the
  contributions of the strange baryon resonance $\Lambda(1405)$ and
  the baryon background with quantum numbers of resonances
  $\Lambda(1800)$ and $\Sigma(1750)$\,\footnote{See ``Comments on the
    approach''.}.  According to Gell--Mann's $SU(3)$ classification of
  hadrons, the $\Lambda(1405)$ resonance is an $SU(3)$ singlet,
  whereas the resonances $\Lambda(1800)$ and $\Sigma(1750)$ are
  components of an $SU(3)$ octet \cite{PDG04}. This allows to describe
  the experimental data on the cross sections for the inelastic
  reactions $K^-p \to Y\pi$, where $Y\pi = \Sigma^-\pi^+$,
  $\Sigma^+\pi^-$, $\Sigma^0\pi^0$ and $\Lambda^0\pi^0$, at threshold
  of $K^-p$ scattering \cite{DT71,RN78}
\begin{eqnarray}\label{label6}
\gamma &=& \frac{\sigma(K^-p \to \Sigma^-\pi^+)}{\sigma(K^-p \to
\Sigma^+\pi^-)} = 2.360 \pm 0.040,\nonumber\\ R_c &=&
\frac{\sigma(K^-p \to \Sigma^-\pi^+) + \sigma(K^-p \to
\Sigma^+\pi^-)}{\sigma(K^-p \to \Sigma^-\pi^+) + \sigma(K^-p \to
\Sigma^+\pi^-) + \sigma(K^-p \to \Sigma^0\pi^0) + \sigma(K^-p \to
\Lambda^0\pi^0)} = \nonumber\\ &=&0.664 \pm 0.011,\nonumber\\ R_n &=&
\frac{\sigma(K^-p \to \Lambda^0\pi^0)}{\sigma(K^-p \to \Sigma^0\pi^0)
+ \sigma(K^-p \to \Lambda^0\pi^0)} = 0.189 \pm 0.015
\end{eqnarray}
with an accuracy of about $6\,\%$ \cite{IV3} and the constraint that
the $\Lambda(1800)$ resonance decouples from the $K^-p$ system at low
energies.  We would like to emphasize that for the description of the
experimental data Eq.(\ref{label6}) we have not specified the values
of the parameters of resonances. We have satisfied the experimental
data by using only the $SU(3)$ properties of the resonances
$\Lambda(1405)$, $\Lambda(1800)$ and $\Sigma(1750)$$^1$.

Due to the optical theorem and using the experimental data
Eq.(\ref{label6}) the imaginary part ${\cal I}m\,f^{K^-p}_0(0)$ of the
S--wave amplitude of $K^-p$ scattering can be expressed in terms of
the parameters $\gamma$ and $R_c$ and the S--wave amplitude $f(K^-p
\to \Sigma^-\pi^+)$ of the inelastic channel $K^-p \to \Sigma^-\pi^+$
\cite{IV3}:
\begin{eqnarray}\label{label7}
{\cal I}m\,f^{K^-p}_0(0) &=& \frac{1}{R_c}\,\Big(1 +
\frac{1}{\gamma}\Big)\,|f(K^-p \to \Sigma^-\pi^+)|^2
k_{\Sigma^-\pi^+} = (0.269\pm 0.032)\,{\rm fm},
\end{eqnarray}
where $k_{\Sigma^-\pi^+} = 173.85\,{\rm MeV}$ is a relative momentum
of the $\Sigma^-\pi^+$ pair at threshold of $K^-p$ scattering.

In our model for the calculation of the amplitude $f(K^-p \to
\Sigma^-\pi^+)$ one needs two input parameters $g_1$ and $g_2$, the
coupling constants of the interactions $\Lambda(1405)$ and
$\Sigma(1750)$ with $B = (N,\Sigma,\Lambda^0,\Xi)$ and $P = (\pi,
K,\bar{K},\eta(550))$, the octets of low--lying baryons and
pseudoscalar mesons. The coupling constants $g_1 = 0.907$ and $g_2 =
1.123$ have been calculated in \cite{IV3}, using the recommended
values for masses and widths of the resonances $\Lambda(1405)$ and
$\Sigma(1750)$ and physical masses of baryons and pseudoscalar mesons
\cite{PDG04}.

The real part ${\cal R}e\,f^{K^-p}_0(0)$ of the S--wave amplitude of
$K^-p$ scattering at threshold, i.e. the S--wave scattering length
${\cal R}e\,f^{K^-p}_0(0) = a^{K^-p}_0$ of $K^-p$ scattering,
\begin{eqnarray}\label{label8}
{\cal R}e\,f^{K^-p}_0(0) = {\cal R}e\,f^{K^-p}_0(0)_R
+ {\cal R}e\,\tilde{f}^{K^-p}_0(0),
\end{eqnarray}
is defined by the contribution (i) of the strange baryon resonances in
the s--channel of low--energy elastic $K^-p$ scattering, which we
denote as ${\cal R}e\,f^{K^-p}_0(0)_R$, (ii) of the exotic four--quark
(or $K\bar{K}$ molecules) scalar states $a_0(980)$ and $f_0(980)$ in
the $t$--channel of low--energy elastic $K^- p$ scattering and (iii)
of hadrons with non--exotic quark structures, i.e.  $q\bar{q}$ for
mesons and $qqq$ for baryons \cite{PDG04}, where $q = u, d$ or $s$
quarks, which we denote as ${\cal R}e\,\tilde{f}^{K^-p}_0(0)$.

The contribution of the strange resonances ${\cal R}e\,f^{K^-p}_0(0)_R
= (- 0.154 \pm 0.009)\,{\rm fm}$  explains $32\,\%$ of the mean value
of the S--wave scattering length of $K^-p$ scattering \cite{IV3}:
${\cal R}e\,f^{K^-p}_0(0)= (-\,0.482 \pm 0.034)\,{\rm fm}$.

The contribution of the exotic scalar mesons $a_0(980)$ and $f_0(980)$
and non--exotic hadrons, calculated to leading order in chiral
expansion, is equal to \cite{IV3}
\begin{eqnarray}\label{label9}
{\cal R}e\,\tilde{f}^{K^-p}_0(0) = 0.398 - 0.614\,\xi\,{\rm fm},
\end{eqnarray}
where the first term is defined by strong low--energy interactions of
non--exotic hadrons, whereas the second one is caused by the exotic
scalar mesons $a_0(980)$ and $f_0(980)$ coupled to the $K^-$--meson
and the proton. The parameter $\xi$ is a low--energy constant (LEC),
defining low--energy $a_0(980)NN$ and $f_0(980)NN$ interactions. 

Using the hypothesis of quark--hadron duality, formulated by
Shifman, Veinshtein and Zakharov within non--perturbative QCD in the
form of the QCD sum rules \cite{VZ79}, and the Effective quark model
with chiral $U(3)\times U(3)$ symmetry \cite{AI99}, we have calculated
the parameter $\xi$ \cite{IV3}: $\xi = 1.2\pm 0.1$.

The validity of the application of the Effective quark model with
chiral $U(3)\times U(3)$ symmetry \cite{AI99} to the quantitative
analysis of strong low--energy $\bar{K}N$ interactions at threshold
has been corroborated by example of the calculation of the S--wave
scattering length of elastic $K^-n$ scattering in connection with the
calculation of the energy level displacement of the ground state of
kaonic deuterium \cite{IV4}. 

For the description of the contributions of the strange baryon
resonances $\Lambda(1405)$, $\Lambda(1800)$ and $\Sigma(1750)$ and the
exotic scalar mesons $a_0(980)$ and $f_0(980)$ we have used Effective
Chiral Lagrangians with $\Lambda(1405)$, $\Lambda(1800)$,
$\Sigma(1750)$, $a_0(980)$ and $f_0(980)$ treated as elementary
particles and represented by local interpolating fields.  Such a
description of baryon resonances and exotic mesons does not contradict
ChPT by Gasser and Leutwyler \cite{JG83} and power counting
\cite{MR96}--\cite{UM01}. As has been pointed out in \cite{MR96},
the inclusion of the resonance $\Lambda(1405)$ as well as
$\Lambda(1800)$ and $\Sigma(1750)$ as elementary particle fields
allows to calculate the amplitudes of $\bar{K}N$ scattering at threshold
to leading order Effective Chiral Lagrangians. 

\subsection{Energy level displacement of the excited $nP$ states 
of kaonic hydrogen}

The application of our model of strong low--energy $\bar{K}N$
interactions at threshold \cite{IV3,IV4} to the description of $K^-p$
scattering in the P--wave state has led to the following value of the
energy level displacement of the excited $nP$ states of kaonic
hydrogen \cite{IV5}
\begin{eqnarray}\label{label10}
  \hspace{-0.3in}\epsilon_{np} &=& -\,\frac{2}{3}\,\frac{\alpha^5}{n^3}\,
\Big(1 - \frac{1}{n^2}\Big)\,
  \Big(\frac{m_Km_N}{m_K + m_N}\Big)^4\,
  (2\,a^{K^-p}_{3/2} +
  a^{K^-p}_{1/2}) = \frac{32}{3}\,\frac{1}{n^3}\,
\Big(1 - \frac{1}{n^2}\Big)\,\epsilon_{2p},\nonumber\\
  \hspace{-0.3in} \Gamma_{np} &=& \frac{4}{9}\,\frac{\alpha^5}{n^3}\,
\Big(1 - \frac{1}{n^2}\Big)\,
  \Big(\frac{m_Km_N}{m_K + m_N}\Big)^4
  \sum_{Y\pi}(2\,a^{Y\pi}_{3/2} +
  a^{Y\pi}_{1/2})^2\,k^3_{Y\pi} = \frac{32}{3}\,\frac{1}{n^3}\,
\Big(1 - \frac{1}{n^2}\Big)\,\Gamma_{2p}
\end{eqnarray}
with $\epsilon_{2p} = -\,0.6\,{\rm meV}$ and $\Gamma_{2p} = 2\,{\rm
  meV}$.

Faifman and Men'shikov have presented the calculated yields for the
$K$--series of $X$--rays for kaonic hydrogen in dependence of the
hydrogen density \cite{MF05}.  They have shown that the use of the
theoretical value $\Gamma_{2p} = 2\,{\rm meV}$ of the width of the
$2p$ state of kaonic hydrogen, calculated in our model, leads to good
agreement with the experimental data, measured for the
$K_{\alpha}$--line by the KEK Collaboration \cite{KEK}.  They have
also shown that the results of cascade calculations with other values
of the width of the $2p$ excited state of kaonic hydrogen, used as an
input parameter, disagree with the available experimental data. The
results obtained by Faifman and Men'shikov contradict to those by
Jensen and Markushin \cite{XR7}.  Therefore, as has been accentuated
by Faifman and Men'shikov \cite{MF05}, the further analysis of the
experimental data by the DEAR Collaboration should allow to perform a
more detailed comparison of the theoretical value $\Gamma_{2p} =
2\,{\rm meV}$ with other phenomenological values of the width of the
$2p$ state of kaonic hydrogen $\Gamma_{2p}$, used as input parameters.

\subsection{Radiative transitions $nP \to 1S + \gamma$ in kaonic
  hydrogen and kaonic deuterium, caused by strong low--energy
  interactions}

The analysis of the transitions $(K^-p)_{np} \to (K^-p)_{1s} + \gamma$,
induced by strong low--energy interactions and enhanced by the Coulomb
interaction of the $K^-p$ pair in the initial and final state of the
amplitude of the kaon--proton {\it Bremsstrahlung} $K^-+ p \to K^- + p
+ \gamma$, has led to the transition rate \cite{IV6}:
\begin{eqnarray}\label{label11}
  \Gamma((K^-p)_{np} \to (K^-p)_{1s}\, \gamma) = \frac{8}{n^3}\,
\frac{\xi^2_{np}}{\xi^2_{2p}}\,\Gamma((K^-p)_{2p} \to (K^-p)_{1s}\, \gamma),
\end{eqnarray}
where the transition rate $\Gamma((K^-p)_{2p} \to (K^-p)_{1s}\,
\gamma)$ is defined in terms of the energy level displacement of the
ground state of kaonic hydrogen:
\begin{eqnarray}\label{label12}
  \Gamma((K^-p)_{2p} \to (K^-p)_{1s}\, \gamma)= 
  \frac{\xi^2_{1s}\xi^2_{2p}}{36}\,
  \frac{\alpha}{\mu}\,\Big(\epsilon^2_{1s} + \frac{1}{4}\,
  \Gamma^2_{1s}\Big) = 4.3\times 10^4
  \,\Big(\epsilon^2_{1s} + \frac{1}{4}\,\Gamma^2_{1s}\Big)\,{\rm sec^{-1}}. 
\end{eqnarray}
The parameters $\xi_{1s}$ and $\xi_{np}\,(n = 2,3,\ldots)$ are the
overlap integrals of the Coulomb wave functions of the bound and
scattering states of the $K^-p$ pairs \cite{IV6}.

For kaonic deuterium we have obtained the following transition rate
\cite{IV6}:
\begin{eqnarray}\label{label13}
  \Gamma((K^-d)_{2p} \to (K^-d)_{1s}\,\gamma) = 3.6\times 10^4\,
\Big(\epsilon^2_{1s} + \frac{1}{4}\,
  \Gamma^2_{1s}\Big)\,{\rm sec^{-1}}.
\end{eqnarray}
Measurements of the energy level displacements of the ground states
for kaonic atoms at the ${\rm eV}$ level would demand to take into
account the transition rates $(K^-p)_{np} \to (K^-p)_{1s} + \gamma$
and $(K^-d)_{np} \to (K^-d)_{1s} + \gamma$, given by
Eqs.(\ref{label12}) and (\ref{label13}), for the theoretical
description of the experimental data on the $X$--ray spectra and
yields.

\subsection{Isospin--breaking and dispersive 
corrections to the 
energy level displacement of the ground state of kaonic hydrogen}

In \cite{IV7} we have applied our model of strong low--energy
$\bar{K}N$ interactions at threshold to the analysis of
isospin--breaking corrections to the S--wave amplitude of $K^-p$
scattering \cite{IV7}, caused by the QCD isospin--breaking interaction
\cite{JG75,JG82}
\begin{eqnarray}\label{label14} 
{\cal L}^{\Delta I = 1}_{\rm \,QCD}(x) =
  \frac{1}{2}\,(m_d - m_u)\,[\bar{u}(x)u(x) - \bar{d}(x)d(x)].
\end{eqnarray}
We have shown that these corrections are of order $O(\alpha)$ and make
up about $0.04\,\%$.  Our results agree well with the estimates,
obtained by Mei\ss ner {\it et al.} \cite{UM04} within a systematic
analysis of isospin--breaking corrections in the framework of a
non--relativistic Effective Field Theory based on ChPT by Gasser and
Leutwyler \cite{JG83}. 

In our model of strong low--energy $\bar{K}N$ interactions we have
derived the formula for the energy level displacement of the ground
state of kaonic hydrogen \cite{IV3,IV4,IV5,IV6,IV7}
\begin{eqnarray}\label{label15}
  - \,\epsilon_{1s} + i\,\frac{\Gamma_{1s}}{2} &=&
  \frac{1}{4m_Km_p}\,\frac{1}{2}\sum_{\sigma_p = \pm 1/2}
\int\frac{d^3k}{(2\pi)^3}
  \int\frac{d^3q}{(2\pi)^3}\,
  \sqrt{\frac{m_K m_p}{E_K(\vec{k}\,)E_p(\vec{k}\,)}}\,
  \sqrt{\frac{m_K m_p}{E_K(\vec{q}\,)
      E_p(\vec{q}\,)}}\nonumber\\ &\times&\Phi^{\dagger}_{1s}(\vec{k}\,)\,
  M(K^-(\vec{q}\,)p(-\vec{q},\sigma_p) \to
  K^-(\vec{k}\,)p(-\vec{k},\sigma_p))\, \Phi_{1s}(\vec{q}\,),
\end{eqnarray}
where $M(K^-(\vec{q}\,)p(-\vec{q},\sigma_p) \to
K^-(\vec{k}\,)p(-\vec{k},\sigma_p))$ is the S--wave amplitude of
$K^-p$ scattering, $\Phi^{\dagger}_{1s}(\vec{k}\,)$ and
$\Phi_{1s}(\vec{q}\,)$ are the wave functions of the ground state of
kaonic hydrogen in the momentum representation normalized to unity: $
\Phi_{1s}(\vec{p}\,) = 8\sqrt{\pi a^3_B}/(1 + p^2 a^2_B)^2$; $a_B =
1/\alpha \mu = 83.6\,{\rm fm}$ is the Bohr radius. The formula for the
energy level displacements of the excited $n\ell$ states of kaonic
hydrogen analogous to Eq.(\ref{label15}) has been derived in
\cite{IV5}.

Inserting the intermediate $\bar{K}^0n$ state on--mass shell, i.e.
taking into account the transition $K^-p \to \bar{K}^0n \to K^-p$, and
keeping the observable mass differences of the neutral and charged $K$
mesons and the neutron and the proton, we have calculated the
dispersive corrections to the energy level displacement of the ground
state of kaonic hydrogen  \cite{IV7}
\begin{eqnarray}\label{label16}
  \hspace{-0.3in}\delta^{Disp}_S &=& \frac{\delta 
    \epsilon^{\bar{K}^0n}_{1s}}{\epsilon^{(0)}_{1s}} = \frac{1}{4}\,
  (a^1_0 - a^0_0)^2\,q^2_0
  = (4.8 \pm 0.4)\,\%,\nonumber\\
  \hspace{-0.3in}\delta^{Disp}_W &=& \frac{\delta 
    \Gamma^{\bar{K}^0n}_{1s}}{\Gamma^{(0)}_{1s}} = 
  \frac{1}{2\pi}\,
  \frac{(a^1_0 -a^0_0)^2}{{\cal I}m\,f^{K^-p}_0(0)\,a_B}\,
  {\ell n}\Big[\frac{2 a_B}{|a^0_0 + a^1_0|}\Big] = (8.0 \pm 1.0)\,\%,
\end{eqnarray}
where $q_0 = \sqrt{2\mu(m_{\bar{K}^0} - m_K + m_n - m_p)} = 58.4\,{\rm
  MeV}$ \cite{PDG04} and ${\cal I}m\,f^{K^-p}_0(0) = (0.269 \pm
0.032)\,{\rm fm}$ \cite{IV3}. The parameter $2/|a^0_0 + a^1_0| = (409
\pm 29)\,{\rm MeV}$, appearing in the argument of the logarithm ${\ell
  n}(2a_B/|a^0_0 + a^1_0|)$ in Eq.(\ref{label16}), has a meaning of a
natural cut--off, introduced by the final--state $\bar{K}^0n$
interaction \cite{IV7}. Such corrections have not been calculated before in
literature.

\subsection{ Energy level shift of the ground state of kaonic hydrogen
  and the $\sigma^{I = 1}_{KN}(0)$--term}

In our approach the S--wave amplitude of $K^-p$ scattering is
calculated to leading order in chiral expansion \cite{IV3}. This
allows to take into account next--to--leading corrections in chiral
expansions such as the $\sigma^{(I = 1)}_{KN}(0)$--term, defined by
\cite{ER72}--\cite{JG99}
\begin{eqnarray}\label{label17}
  \sigma^{(I = 1)}_{KN}(0) = \frac{m_u + m_s}{4m_N}\,
\langle p(\vec{0},\sigma_p)|\bar{u}(0)u(0) + \bar{s}(0)s(0)|p(\vec{0},
\sigma_p)\rangle,
\end{eqnarray}
where $m_u$ and $m_s$ are masses and $u(0)$ and $s(0)$ are operators
of the interpolating fields of $u$ and $s$ current quarks.

The contribution of the $\sigma^{(I = 1)}_{KN}(0)$--term to the energy
level shift is equal to \cite{IV5,IV7}
\begin{eqnarray}\label{label18}
   \delta \epsilon^{(\sigma)}_{1s} =  \frac{\alpha^3 \mu^3}{2\pi m_K 
F^2_K}\Big[\sigma^{(I = 1)}_{KN}(0) - 
  \frac{m^2_K}{4m_N}i\int d^4x\langle p(\vec{0},\sigma_p)
  |{\rm T}(J^{4+i5}_{50}(x)J^{4-i5}_{50}(0))|
  p(\vec{0},\sigma_p)\rangle\Big],
\end{eqnarray} 
where $J^{4\pm i5}_{50}(x)$ are time--components of the axial hadronic
currents $J^{4\pm i5}_{5\mu}(x)$, changing strangeness $|\Delta S| =
1$, $F_K = 113\,{\rm MeV}$ is the PCAC constant of the $K$--meson
\cite{PDG04}.

The correction $\delta \epsilon^{(\sigma)}_{1s}$ to the shift of the
energy level of the ground state of kaonic hydrogen, caused by the
$\sigma^{(I = 1)}_{KN}(0)$, is obtained from the S--wave amplitude of
$K^-p$ scattering, calculated to next--to--leading order in ChPT
expansion at the tree--hadron level \cite{IV5} and Current Algebra
\cite{ER72}:
\begin{eqnarray}\label{label19}
 \hspace{-0.3in} &&4\pi\,\Big(1 + \frac{m_K}{m_N}\Big)\,f^{K^-p}_0(0) = 
\frac{m_K}{F^2_K} - 
  \frac{1}{F^2_K}\,\sigma^{(I = 1)}_{KN}(0)\nonumber\\
 \hspace{-0.3in}  &&\hspace{1in} + 
  \frac{m^2_K}{4m_N F^2_K}\,i\int d^4x\,\langle p(\vec{0},\sigma_p)
  |{\rm T}(J^{4+i5}_{50}(x)J^{4-i5}_{50}(0))|
  p(\vec{0},\sigma_p)\rangle.
\end{eqnarray} 
Since the first term, $m_K/F^2_K$, calculated to leading order in
chiral expansion, has been already taken into account in \cite{IV3},
the second term, $- \sigma^{(I = 1)}_{KN}(0)/F^2_K$, and the third one
define next--to--leading order corrections in chiral expansion to the
S--wave amplitude of $K^-p$ scattering at threshold.

Taking into account the contribution $\delta
\epsilon^{(\sigma)}_{1s}$, the total shift of the energy level of the
ground state of kaonic hydrogen is equal to \cite{IV5,IV7}:
\begin{eqnarray}\label{label20}
  \hspace{-0.3in} \epsilon^{(\rm th)}_{1s} &=&  (213 \pm 15) +
  \frac{\alpha^3 \mu^3}{2\pi m_K 
    F^2_K}\,\sigma^{(I = 1)}_{KN}(0)\nonumber\\
  \hspace{-0.3in} &-& \frac{\alpha^3 \mu^3 m_K}{ 
    8\pi  F^2_K m_N}\,i\int d^4x\,\langle p(\vec{0},\sigma_p)
  |{\rm T}(J^{4+i5}_{50}(x)J^{4-i5}_{50}(0))|
  p(\vec{0},\sigma_p)\rangle.
\end{eqnarray}
The theoretical estimates of the value of $\sigma^{(I = 1)}_{KN}(0)$,
carried out within Effective Field Theory approach based on ChPT with
a dimensional regularization of divergent integrals, are converged
around the number $\sigma^{(I = 1)}_{KN}(0) = (200 \pm 50)\,{\rm MeV}$
\cite{VB93,BB99}.  Hence, the contribution of $\sigma^{(I =
  1)}_{KN}(0)$ to the energy level shift amounts to
\begin{eqnarray}\label{label21}
 \frac{\alpha^3 \mu^3}{2\pi m_K 
    F^2_K}\,\sigma^{(I = 1)}_{KN}(0)  =  (67\pm 17)\,{\rm eV}.
\end{eqnarray}
Substituting Eq.(\ref{label21}) into Eq.(\ref{label20}) we obtain the
total shift of the energy level of the ground state of kaonic hydrogen
equal to
\begin{eqnarray}\label{label22}
   \epsilon^{(\rm th)}_{1s} =  (280 \pm 23) - 
  \frac{\alpha^3 \mu^3 m_K}{ 
    8\pi F^2_K m_N}\,i\int d^4x\,\langle p(\vec{0},\sigma_p)
  |{\rm T}(J^{4+i5}_{50}(x)J^{4-i5}_{50}(0))|
  p(\vec{0},\sigma_p)\rangle.
\end{eqnarray}
The theoretical analysis of the second term in Eq.(\ref{label22}) is
required for the correct understanding of the contribution of the
$\sigma^{I = 1}_{KN}(0)$--term to the shift of the energy level of the
ground state of kaonic hydrogen.

Of course, one can solve the inverse problem. Indeed, carrying
out a theoretical calculation of the term 
\begin{eqnarray}\label{label23}
  \frac{\alpha^3 \mu^3 m_K}{ 
    8\pi F^2_K m_N}\,i\int d^4x\,\langle p(\vec{0},\sigma_p)
  |{\rm T}(J^{4+i5}_{50}(x)J^{4-i5}_{50}(0))|
  p(\vec{0},\sigma_p)\rangle
\end{eqnarray}
in Eq.(\ref{label20}) and using the experimental data on the shift of
the energy level of the ground state of kaonic hydrogen, measured by
the DEAR Collaboration, one can extract the value of the $\sigma^{(I =
  1)}_{KN}(0)$--term. Such a value of the $\sigma^{(I =
  1)}_{KN}(0)$--term should be compared with the theoretical
predictions, obtained within the Effective Field Theory approach based
on ChPT by Gasser and Leutwyler.

\section{Kaonic deuterium}

The energy level displacement of the ground state of kaonic deuterium
is given by the DGBTT formula \cite{DT54} (see also \cite{IV4}):
\begin{eqnarray}\label{labe24}
- \epsilon_{1s} + i\,\frac{\Gamma_{1s}}{2} = 2\alpha^3\mu^2\,
f^{K^-d}_0(0) = 602\,f^{K^-d}_0(0),
\end{eqnarray}
where $\mu = m_Km_d/(m_k + m_d) = 391\,{\rm MeV}$ is the reduced mass
of the $K^-d$ pair for $m_K = 494\,{\rm MeV}$ and $m_d = 1876\,{\rm
  MeV}$ and $f^{K^-d}_0(0)$ is the S--wave amplitude of $K^-d$
scattering at threshold.

In our model of strong low--energy $\bar{K}N$ interactions \cite{IV4}
we have shown that the S--wave scattering length ${\cal
  R}e\,f^{K^-d}_0(0) = a^{K^-d}_0$ of $K^-d$ scattering is fully
defined by the Ericson--Weise scattering length for $K^-d$ scattering
in the S--wave state \cite{TE88}, which we denote as
$(a^{K^-d}_0)_{\rm EW}$, i.e. $a^{K^-d}_0 \simeq (a^{K^-d}_0)_{\rm
  EW}$. The S--wave scattering length $(a^{K^-d}_0)_{\rm EW}$ is equal to
\cite{IV4}
\begin{eqnarray}\label{label25}
  (a^{K^-d}_0)_{\rm EW} &=& \frac{1 + m_K/m_N}{1 + m_K/m_d}\,(a^{K^-p}_0 +
  a^{K^-n}_0)\nonumber\\
  &+& \frac{1}{4}\,\Big(1 +
  \frac{m_K}{m_d}\Big)^{-1}\Big(1 + \frac{m_K}{m_N}\Big)^2\,((a^1_0)^2 +
  4 a^0_0 a^1_0 - (a^0_0)^2)\, \Big\langle \frac{1}{r_{12}}\Big\rangle,
\end{eqnarray}
where $r_{12}$ is a distance between two scatterers $n$ and $p$
\cite{TE88}. In our approach $\langle 1/r_{12}\rangle$ is defined by
\cite{IV4}
\begin{eqnarray}\label{label26}
\Big\langle \frac{1}{r_{12}}\Big\rangle = \int
d^3x\,\Psi^*_d(\vec{r}\,)\,\frac{\displaystyle e^{\textstyle\,- m_K
r}}{r}\,\Psi_d(\vec{r}\,) = 2\gamma_d \,E_1\Big(\frac{m_N}{m_K
+ 2 \gamma_d}\Big) = 0.29\,m_{\pi},
\end{eqnarray}
where $\Psi_d(\vec{r}\,)$ is the wave function of the deuteron in the
ground state. 

We would like to remind that Ericson and Weise did not investigate the
S--wave scattering length of $K^-d$ scattering. They analysed only
$\pi^-d$ scattering. However, since the structure of the term
Eq.(\ref{label25}) is very similar to that of $\pi^-d$ scattering we
call such a term as the Ericson--Weise scattering length
$(a^{K^-d}_0)_{\rm EW}$.  At the quantum field theoretic
level the Ericson--Weise scattering length has been derived in
\cite{IV4}.

The imaginary part ${\cal I}m\,f^{K^-d}_0(0)$ of the S--wave amplitude
of $K^-d$ scattering at threshold we have calculated in terms of the
contributions of the two--body reactions $K^-d \to NY$, where $NY =
n\Lambda^0$, $n\Sigma^0$ and $p\Sigma^-$, and the experimental data on
the two--body production rates \cite{VV70}.  Our theoretical
predictions for the ratios of the two--body production rates \cite{IV4}
\begin{eqnarray}\label{label27}
R(\Lambda^0/\Sigma^0) &=& \frac{{\cal
I}m\,\tilde{f}^{\,K^-d}_0(0)_{n\Lambda^0}}{{\cal
I}m\,\tilde{f}^{\,K^-d}_0(0)_{n\Sigma^0}} = 1.0 \pm 0.3,\nonumber\\
R(\Sigma^0/\Sigma^-) &=& \frac{{\cal
I}m\,\tilde{f}^{\,K^-d}_0(0)_{n\Sigma^0}}{{\cal
I}m\,\tilde{f}^{\,K^-d}_0(0)_{p\Sigma^-}} = 0.8 \pm 0.2, \nonumber\\
R(\Lambda^0/\Sigma^-) &=& \frac{{\cal
I}m\,\tilde{f}^{\,K^-d}_0(0)_{n\Lambda^0}}{{\cal
I}m\,\tilde{f}^{\,K^-d}_0(0)_{p\Sigma^-}} = 0.8 \pm 0.2
\end{eqnarray}
agree well with the experimental data \cite{VV70}
\begin{eqnarray}\label{label28}
R(\Lambda^0/\Sigma^0) &=& \frac{R(K^-d \to n \Lambda^0)}{R(K^-d \to n
\Sigma^0)} = (1.15 \pm 0.27),\nonumber\\
R(\Sigma^0/\Sigma^-) &=&
\frac{R(K^-d \to n \Sigma^0)}{ R(K^-d \to p \Sigma^-)} = (0.68 \pm
0.15),\nonumber\\ R(\Lambda^0/\Sigma^-) &=& \frac{R(K^-d \to n
\Lambda^0)}{ R(K^-d \to p \Sigma^-)} = (0.77 \pm 0.10).
\end{eqnarray}
The experimental data on the production rates $R(K^-d \to NY)$ are
\cite{VV70}
\begin{eqnarray}\label{label29}
R(K^-d \to n \Lambda^0) &=& (0.387 \pm 0.041)\,\%,\nonumber\\ R(K^-d
\to n \Sigma^0) &=& (0.337 \pm 0.070)\,\%,\nonumber\\ R(K^-d \to p
\Sigma^-) &=& (0.505 \pm 0.036)\,\%,\nonumber\\ R = \sum_{NY} R(K^-d
\to NY) &=& (1.229 \pm 0.090)\,\%.
\end{eqnarray}
The theoretical values of the partial widths of two--body inelastic
decay channels $K^-d \to NY$, obtained in our model, are equal to \cite{IV4}
\begin{eqnarray}\label{label30}
\Gamma^{(n \Lambda^0)}_{1s} = (2.2 \pm 0.5)\,{\rm eV},
\Gamma^{(n \Sigma^0)}_{1s} = (2.4 \pm 0.5)\,{\rm eV},
\Gamma^{(p \Sigma^-)}_{1s} &=& (3.0 \pm 0.6)\,{\rm eV}.
\end{eqnarray}
Using the experimental value of the total production rate $R =
(1.229\pm 0.090)\,\%$ and our theoretical predictions for the widths
of the two--body decay channels (\ref{label30}), we can estimate the
expected value of the total width of the ground state of kaonic
deuterium
\begin{eqnarray}\label{label31}
\Gamma_{1s} = \frac{\Gamma^{(n\Lambda^0)}_{1s} + \Gamma^{(n
\Sigma^0)}_{1s} + \Gamma^{(p \Sigma^-)}_{1s}}{(1.229 \pm 0.090)\times
10^{-2}} = (630 \pm 100)\,{\rm eV}.
\end{eqnarray}
The S-wave amplitude of $K^-d$ scattering at
threshold is equal to \cite{IV4}:
\begin{eqnarray}\label{label32}
f^{K^-d}_0(0) = (-\,0.540 \pm 0.095) +
i\,(0.521 \pm 0.075)\, {\rm fm}
\end{eqnarray}
Our prediction for the energy level displacement of the ground state
of kaonic deuterium  \cite{IV4}
\begin{eqnarray}\label{label33}
-\,\epsilon_{1s} + i\,\frac{\Gamma_{1s}}{2} = 602\,f^{K^-d}_0(0) = 
 (-\,325 \pm 60) + i\,(315 \pm 50)\;{\rm eV}.
\end{eqnarray}
can be used for the planning experiments by the DEAR/SIDDHARTA
Collaborations at Frascati \cite{SIDDHARTA}.

\section{Perspectives}

Within our approach to the description of strong low--energy $\bar{K}N$
interactions one can solve the following problems :

\begin{itemize}
\item The extraction of the value of the $\sigma^{I =
    1}_{KN}$--term from the experimental data by the DEAR Collaboration
  on the energy level displacement of kaonic hydrogen
\item The calculation of the energy level displacement of the ground
  state of kaonic deuterium, caused by three--body inelastic channels
  $K^-d \to NY\pi$
\item The analysis of the contribution of the reaction $K^-d
    \to \bar{K}^0nn \to K^-d$ to the energy level displacement of the
    ground state of kaonic deuterium
  \item The analysis of the contribution of the $\sigma^{I =
      1}_{KN}$--term to the shift of the energy level of the ground
    state of kaonic deuterium
\item The calculation of the energy level displacement of the
    excited $nP$ states of kaonic deuterium
\item The calculation of the energy level displacement of the
    ground state and $nP$ excited states of kaonic ${^3}{\rm
      He}$ and ${^4}{\rm He}$
  \item The calculation of the radiative transition rates $nP \to 1S +
    \gamma$ for kaonic ${^3}{\rm He}$ and ${^4}{\rm He}$, induced by
    strong low--energy interactions and enhanced by the Coulomb
    interaction in the $K^-{^3}{\rm He}$ and $K^-{^4}{\rm He}$ pairs
  \item The analysis of the antiprotonic atoms and the deeply bound
    states $(K^-pp)$ and $(K^-NNN)$.
\end{itemize}

\section{Comments on the approach}

\subsection{Kaonic hydrogen revisited}

In our recent analysis of the S--wave scattering lengths $a^{K^-0}_0 =
(a^0_0 + a^1_0)/2$ and $a^{K^-n}_0 = a^1_0$ of $K^-p$ and $K^-n$
scattering, carried out in \cite{IV8}, we have shown that the S--wave
scattering lengths $a^{K^-0}_0$ and $a^{K^-n}_0$, calculated to
leading order in chiral expansion, satisfy the low--energy theorem
\begin{eqnarray}\label{label34}
a^{K^-p}_0 + a^{K^-n}_0 = \frac{1}{2}\,(a^0_0 + 3\,a^1_0) = 0.
\end{eqnarray}
We have derived this low--energy theorem relating the S--wave
scattering lengths of $K^-N$ scattering to the S--wave scattering
lengths of $\pi^- N$ scattering.

As has been shown by Weinberg \cite{SW66}, in the chiral limit the
S--wave scattering lengths $a^{\pi^-p}_0 = (2\,a^{1/2}_0 +
a^{3/2}_0)/3$ and $a^{\pi^-n}_0 = a^{3/2}_0$ of $\pi^-N$ elastic
scattering\,\footnote{Here $a^{1/2}_0$ and $a^{3/2}_0$ are the S--wave
scattering lengths of $\pi N$ scattering with isospin $I = 1/2$ and $I
= 3/2$.} obey the constraint
\begin{eqnarray}\label{label35}
  a^{\pi^-p}_0 + a^{\pi^-n}_0 = 
  \frac{2}{3}\,(a^{1/2}_0 + 2a^{3/2}) = 2\,b^0_0 = 0,
\end{eqnarray}
which is caused by Adler's consistency condition
\cite{SA65}\,\footnote{We remind that $b^0_0$ is the S--wave
  scattering length of $\pi N$ scattering in the $t$--channel with
  isospin $I = 0$.}. We have shown \cite{IV8} that due to isospin
  invariance of strong low--energy interactions in the chiral limit
  the sum of the S--wave scattering lengths $a^{K^-p}_0 + a^{K^-n}_0$
  is equal to
\begin{eqnarray}\label{label36}
a^{K^-p}_0 + a^{K^-n}_0 = -\,\sqrt{6}\,b^0_0 = 0.
\end{eqnarray}
The same result can be obtained by using $U$--spin invariance of
strong low--energy interactions in the chiral limit. According to
$U$--spin classification, the $K$ and $\pi$ mesons are components of
$U$--spin doublets.

We have found that the $\Sigma(1750)$ resonance does not saturate the
low--energy theorem Eq.(\ref{label34}), therefore we have replaced the
contribution of the $\Sigma(1750)$ resonance by the phenomenological
baryon background with quantum numbers of the $\Sigma(1750)$ resonance
\cite{IV8}. The value of this background we have fixed in terms of the
parameter $\gamma$ Eq.(\ref{label6}) and the contribution of the
$\Lambda(1405)$ resonance.

For the complex S--wave scattering lengths $\tilde{a}^0_0$ and
$\tilde{a}^1_0$ of $\bar{K}N$ scattering with isospin $I = 0$ and
$I = 1$ we have  obtained the numerical values 
\begin{eqnarray}\label{label37}
  \tilde{a}^0_0 &=& 
  (-\,1.50\pm 0.05)  + \,i(\,0.66 \pm 0.01)\,{\rm fm},
  \nonumber\\
  \tilde{a}^1_0  &=&
  (+\,0.50\pm 0.02)   + \,i\,(0.04 \pm 0.00)\,{\rm fm},
\end{eqnarray}
where ${\cal R}e\,\tilde{a}^0_0 = a^0_0 = (-\,1.50\pm 0.05)\,{\rm
  fm}$ and ${\cal R}e\,\tilde{a}^1_0 = a^1_0 = (+\,0.50\pm
0.02)\,{\rm fm}$.  

It is seen that the real parts of the complex scattering lengths
$a^0_0$ and $a^1_0$ satisfy the low--energy theorem
Eq.(\ref{label34}).

The complex S--wave scattering length $\tilde{a}^0_0$ agrees well with
the result obtained by Dalitz and Deloff \cite{RD91}
\begin{eqnarray}\label{label38}
\tilde{a}^0_0 = (-\,1.54 \pm 0.05) + \,i\,(0.74 \pm 0.02)\,{\rm fm}
\end{eqnarray}
\noindent for the position of the pole on sheet II of the $E$--plane
$E^* - i\,\Gamma/2$ with $E^* = 1404.9\,{\rm MeV}$ and $\Gamma =
53.1\,{\rm MeV}$ \cite{RD91}. This corresponds to our choice of the
parameters of the $\Lambda(1405)$ resonance.

It is interesting to notice that the complex S--wave scattering length
$\tilde{a}^0_0$, calculated in our model, does not contradict the
result obtained by Akaishi and Yamazaki \cite{YA02} at the assumption
that the $\Lambda(1405)$ resonance is the bound $K^-p$ state.

For the energy level displacement of the ground state of kaonic
hydrogen, calculated for the S--wave amplitude of $K^-p$ scattering at
threshold Eq.(\ref{label37}), we get
\begin{eqnarray}\label{label39}
  - \,\epsilon^{(0)}_{1s} + i\,\frac{\Gamma^{(0)}_{1s}}{2} = 
412\,f^{K^-p}_0(0) = 412\,\frac{\tilde{a}^0_0 + 
\tilde{a}^1_0}{2} = (-\,205 \pm 21) 
+ \,i\,(144 \pm 9)\,{\rm eV}.
\end{eqnarray}
This result agrees well with the experimental data by the DEAR
Collaboration.

We have shown that the cross sections for elastic $K^-p$ scattering
and inelastic channels $K^-p \to \Sigma^-\pi^+$, $K^-p \to
\Sigma^+\pi^-$, $K^-p \to \Sigma^0\pi^0$ and $K^-p \to
\Lambda^0\pi^0$, calculated in our model for laboratory momenta of the
incident $K^-$ meson from domain $70\,{\rm MeV}/c \le p_{lab} \le
150\,{\rm MeV}/c$, satisfy the available experimental data within two
standard deviations. These results agree well with the theoretical
analysis of low--energy $\bar{K}N$ interactions, carried out in
\cite{WW04} within the $SU(3)$ effective chiral Lagrangian approach
and relativistic coupled channels technique.

\subsection{Kaonic deuterium revisited}

Our theoretical predictions for the S--wave amplitude of $K^-d$
scattering and the energy level displacement of the ground state of
kaonic deuterium, given in \cite{IV4}, are left practically unchanged.

\end{document}